\documentclass[12pt]{article}
\usepackage{latexsym}
\usepackage{epsfig}
\usepackage{graphicx}
\usepackage[english]{babel}
\usepackage{epsfig}
\newcommand \Pomeron {I\!\!P}

\input colordvi
\begin{document} 

\title{Large t diffractive $\rho$ meson photoproduction with target dissociation
in ultraperipheral pA and AA collisions at LHC.}

\author {
L. ~Frankfurt\\
\it School of Physics and Astronomy,\\
\it  Tel Aviv University,
Tel Aviv, 69978 , Israel\\
M. ~Strikman\\
\it Department of Physics,
Pennsylvania State University,\\
\it University Park, PA  16802, USA\\
 M.~Zhalov\\
\it St. Petersburg Nuclear Physics Institute, Gatchina, 188300
Russia}
\date{}
\maketitle
\begin{abstract}
We demonstrate  that study of large t vector meson photoproduction 
with rapidity gaps in ultraperipheral proton-ion and ion-ion collisions 
at LHC  would allow  to investigate the energy dependence of cross 
section of elastic scattering of a small dipole off the parton over a 
wide range of energies  $10^3 <  s_{dipole - parton} < 10^6 \, GeV^2$ 
where this cross section is expected to change by a factor $\ge 20$. 
The accessible energy range exceeds  the one reached at HERA by a 
factor of 10 both in $\gamma p$ and in $\gamma A$ scattering.  
In addition,  study of  A-dependence of this process will allow to 
determine the t range where  interaction of small dipoles gives the 
dominant contribution and   to investigate effects of absorption for 
the propagation of ultrarelativistic small $q\bar q$ dipoles through 
the nuclear media and probe in a novel way onset of the black disk regime.
\end{abstract}

\section{Introduction}

One of the principal features of small x processes is a nontrivial
interplay between
the evolution in $\ln(x_o/x)$ and in $\ln(Q^2/Q_o^2)$.  It has been understood
long ago that investigations of large $t$  processes with a large rapidity gap 
give principal possibility to ensure that the coupling constant is small
and the  evolution of the  hard scale is suppressed. As a result such processes allow 
to investigate  $\ln (x_o/x)$ physics separately from the $\ln(Q^2/Q_o^2)$ 
evolution.
Challenging QCD phenomena include transition from the 
regime of nuclear color transparency to color opacity and  
other regimes of high energy strong interaction with a small coupling constant. 
Though the color transparency phenomenon is established
experimentally in \cite{Danny,Sokoloff} further studies are   certainly 
necessary to  establish the range of energies and virtualities for which 
phenomenon takes place. Also, there is a number of indirect indications 
for onset of the regime of color opacity, a direct evidence is very limited, 
see however \cite{FS06}. The rapidity gap processes we discuss in this
paper will provide additional handles to address these questions.

To probe this physics a number of small x processes  
which originate due to elastic scattering of  a 
parton and a small quark-antiquark ($q\bar q$)
color singlet
dipoles (we  will refer to them in the following  simply as dipoles) 
at large momentum transfer and  at high energies were suggested.
 This includes hard diffraction in $pp\to pX$ process at large $t$, production 
of two jets accompanied by rapidity gap-coherent Pomeron \cite{FS89}, 
the rate of production of two back to back jets with a large rapidity gap 
in  between \cite{Bjorken} as compared to the rate of  two jet production 
in the same kinematics without rapidity gaps \cite{MT,DT}, photo(electro) 
production  of vector mesons at large t  with a rapidity gap 
\cite{AFS,FR95,Bartels}. Production of two jets with a gap in between was 
studied experimentally at  the Tevatron, see e.g. \cite{Abe:1998ip}. Over the last ten 
years the theoretical and experimental studies were
focused on the 
photo/electro production off a proton. 
Studies of these processes at 
HERA resulted in the measurements of the relevant cross sections 
\cite{Derrick:1995pb,Adloff:2002em,Chekanov:2002rm,Aktas:2003zi,Aktas:2006qs}
in a region of the photon-proton center of mass energies 
$20\, GeV\le W_{\gamma p}\le 200\, GeV$.
 
The HERA data 
agree well with many (though not all) predictions of the QCD motivated  
models 
(several of which use the LO BFKL approximation\cite{BFKL}),  
see for example \cite{Aktas:2006qs} and references therein.  

Clearly it would be beneficial to 
extend such
study  
to 
higher $W_{\gamma p}$ and over 
a larger range of the rapidity gap intervals to investigate {\it how energy 
dependence of the small dipole - parton scattering changes with t.}
Recently we demonstrated \cite{FSZ06}  that 
this will be possible using
quasireal photons in 
the ultraperipheral collisions (UPC) of protons with nuclei at LHC.

Here we 
perform a more detailed analysis focusing on study of $\rho$ meson
photoproduction: 
\begin{equation}
\gamma + p(A) \to \rho + {\it rapidity\, gap} + X,
\label{eqvm}
\end{equation}
at large 
t and 
with a rapidity gap 
between $\rho$-meson and produced hadronic system $X$
in the proton-nucleus and nucleus-nucleus
UPC at LHC. 
We consider the kinematics where 
 the rapidity gap interval 
is sufficiently large ($\ge 4$) to suppress contribution 
of the fragmentation processes.
 Related physics can be investigated in the 
diffractive production of charm or  two jets separated by large rapidity gap 
from the nucleon fragmentation region. 
For example, studies of the A-dependence 
of production of two jets in the processes like 
$\gamma + A\to (jet + M_1)+ rapidity\, gap +(jet+M_2)$ 
will allow to check presence of the color transparency effects in the gap 
survival in hard photon induced processes \cite{FS96}.
 
The CMS and ATLAS detectors are 
well suited for observing such processes since they cover large rapidity  intervals. 

The main variables determining the dynamics of the process 
are the mass $M_X$ of system produced due to the dissociation of proton target,
the square of the transfered momentum $-t\equiv Q^2= -(p_{\gamma}-p_{V})^2$,  
and the invariant energy of 
the $q\bar q$- parton elastic scattering 
\begin{equation}
s'=xW_{\gamma p}^2,
\end{equation} 
where 
\begin{equation}
x=\frac {-t} {(-t+M_X^2-m_N^2)},
\label{xmin}
\end{equation}
is a fraction of the proton momentum  carried by the parton involved in 
the elastic scattering off the $q\bar q$ pair.
The   rapidity gap between the produced vector meson and knocked out 
parton (roughly corresponding to the leading edge of the rapidity range 
filled by the hadronic system $X$) is 
related to $W_{\gamma p}$ and $t$  (for large $t,W_{\gamma p}^2$) as 
$$y_{r}=\ln \frac {xW_{\gamma p}^2} {\sqrt{(-t)(m_V^2-t)}}.$$
Obviously it is rather difficult  to  measure the mass of diffractively 
produced system $M_X$ or the value of $x$ 
directly.
 However it can be  determined with a good accuracy by measuring momenta 
of the   particles in the system X closest to the rapidity gap.

Generally, the cross sections of the processes of large $t$ scattering with 
a rapidity gap  are given by an incoherent sum of terms describing elastic 
scattering off a quark and off a gluon.
Each of the  terms is
a product
of the cross section of the quasielastic large t  transition
 $projectile \, + \, parton \to final \, hadron \, + \, parton$
 and the corresponding parton density in the  target \cite{FS89,AFS,FR95}. 
The choice of large $t$ ensures
 two important simplifications. First, 
  the   parton ladder
mediating quasielastic  scattering 
 is attached to the  projectile  via two gluons. Second is that  
 attachment of the ladder to two partons of the 
target is strongly suppressed. 
 The elastic cross section  of the scattering off a gluon is enhanced by 
a factor of 81/16 as compared to the scattering off a quark. Numerically, the scattering off the gluons  dominates for a wide range of x. 
For example it constitutes $\sim 80\% (70\%)$  of the cross section at x=0.1(0.3).
The power law for the t-dependence of the cross section is related to the number 
of constituents in the hadronic vertex, hence, 
one can expect 
$1/t^6$ for three quarks \cite{FS89} and 
$1/t^4$ for the quark-antiquark system considered in
\cite{FR95}.
 
In the case of the photoproduction of vector mesons for the $t$
range $-t \gg 1/r_V^2 >  \Lambda_{QCD}^2$ ($r_V$ is the radius of a vector meson),
$W_{\gamma p}^2\gg M_X^2\gg M_N^2$,
 and fixed $x< 1$, 
the target dissociative cross section of the vector meson photoproduction
off proton can be written in the form \cite{FR95}:
\begin{eqnarray}
\frac
{d\sigma_{\gamma+p\to V+X}} {dt dx}=
\frac {d\sigma_{\gamma+quark\to V+quark}} {dt} 
 \biggl [{81\over 16} g_{p}(x,t) 
+\sum_i (q_{p}^{i}(x,t)+{\bar q}_{p}^{i}(x,t)) \biggr ].
\label{DGLAPBFKL}
\end{eqnarray}
Here $g_{p}(x,t)$ is the gluon distribution,  
$q_{p}^{i}(x,t)$ and $\bar q_{p}^{i}(x,t)$ - quark and antiquark 
distributions with flavor $i$ in the proton target. The amplitude of vector meson photoproduction off a quark $f_q (s^\prime ,t)$  is  dominated by a quasielastic scattering of a photon in a small 
quark-antiquark dipole configuration which transits to a quark-antiquark 
component in the vector meson. 

Diffractive photoproduction of vector meson from hadron(nuclear)
target is a particular case of special hard processes where
evolution in $x$ is separated from the evolution in hard scale 
cf. discussions in \cite{FS89,MT,AFS,FR95}. Really, since
$t$ which guarantees smallness of the 
coupling constant is the same
within the ladder mediating quasielastic scattering, the amplitude $f_
{q}(s^{\prime},Q^2=-t)$ probes evolution in $ln(1/x)$ at fixed virtuality
$Q^2$. 
Note also, that since
the the momentum transfer is shared by two gluons 
characteristic virtualities of the t-channel 
gluons in the 
ladder
 are $\approx Q^2/4$. 
At the same time hard scale characterizing
structure function of a target is $\approx Q^2$.

In the lowest order in $\ln(1/x)$   the elastic dipole -quark amplitude, 
$f_{q}(s^{\prime},t)$ ,  is  independent of $W_{\gamma p}$
for any fixed $t$. 
Account of  the higher order terms in $\ln (1/x)$ was performed 
in the leading and next to leading log approximation as well as 
modeling effects of both logs of $Q^2$ and x. It leads to expectation
\footnote{Note however that 
within NLO BFKL this is not a foregone expectation as
with increase of $Q^2$  the solution may be given by a 
different saddle point \cite{Lipatov,Colferai:1999em}.} of 
$f_q$ increasing with energy as a power of $ s^{\prime}/|t|$
which should weakly depend on $t$:
$$f_q(s^{\prime},t)\propto 
\biggl (\frac {s^{\prime}} {|t|}\biggr )^{\delta(t)}.$$
The numerical results  for $\delta(0)$ are strongly modified
when going from LO BFKL: $\delta(0) \sim 0.6$,  to NLO BFKL - $\delta(0) \sim 0.1$, 
and from NLO  to resummed BFKL where  
$\delta(0)\sim 0.2 \div 0.25$ in a wide range of virtualities, 
for the recent review see \cite{Salam:2005yp}. 
Hence we will  treat $\delta(t)$ as a free parameter
and assume, in most of the analysis, 
 that $\delta(t)$ weakly depends on $t$.

Note that similar process for small $t$ could be described by the triple 
Pomeron approximation. In this case 
$f_q \propto (W_{\gamma p}^2/M_X^2)^{\alpha_{\Pomeron}(t)-1}$
where $\alpha_{\Pomeron}(t)\approx 1.08+ 0.25 \,GeV^{-2}t$ is the Pomeron trajectory. 
For $-t \ge 0. 5 \,GeV^{-2}$ this model predicts $f_q$ dropping with increase of energy.

The paper is organized as follows. In section \ref{sec2gap} we analyze the current 
HERA data, and propose a simple parametrization  of the data based on the hard 
mechanism of the reaction. We also estimate the rates 
of $\gamma + p \to \rho + \, gap +\, X$ reaction reaction in $pA $ collisions 
at LHC and find that it will be possible to extend the HERA measurements to 
the energies exceeding those reached at HERA by a factor of 10. 
In section \ref{sec3gap}
we analyze the A-dependence of the discussed process and find it will 
provide a critical test of the interplay between hard and soft dynamics 
as well as probe onset of hard black disk regime. We also find that it 
will be possible to reach $W_{\gamma p}\sim 1\, TeV$ in this process and hence study 
hard dynamics up to  $xW_{\gamma p}^2/Q^2 \sim 10^{5}$ where emission of several gluons in the 
ladder kinematics becomes possible.

\section{Modeling of the elementary and rapidity gap process from HERA to LHC 
in ultraperipheral  $pA$ collisions}
\label{sec2gap}
The current HERA measurements  report cross sections integrated over
$M_X$ from the minimal value $M_X=m_N$ up to some experimentally
fixed  upper limit of ${\hat M}$.  At fixed $t$ this obviously corresponds to
the cross section integrated over $x$ from $x=1$ to the value $x_{min}$ 
determined from Eq.(\ref{xmin}) at $M_{X}={\hat M}$.

Two types of cuts are considered - $x\ge x_{min}$ and $M_X<\hat M$, to reveal 
the dynamics of the process (\ref{eqvm}).  Most of the ZEUS and H1 
measurements explored a rather narrow interval of $W_{\gamma p}$ and were 
mostly focused on the $t$ dependence of the cross section. Evaluation within 
pQCD of  the transition $\gamma \to \rho$ shows that at large $t$ the 
cross section should
be proportional to $|t|^{-4}$.  The H1 collaboration
presented data on the t-dependence of the cross section in  the t-range 
$1\,GeV^2 \le -t\le 8 \,GeV^2$ taken at the average value 
$<W_{\gamma p}>\approx 85 \,GeV$.  They 
measured also energy dependence of 
the cross section for values of $-t<2.5$ GeV$^2$ in the range  
$30 \,GeV\le W_{\gamma p}\le 90\, GeV$ with the cuts on the mass of produced 
system $M_X \le 0.1W_{\gamma p}$. The t-dependence of cross section of the 
diffractive $\rho$, $\phi$ and $J/\psi$ photoproduction with proton dissociation 
was measured by ZEUS at the average value $<W_{\gamma p}>\approx 100$ GeV 
with  restriction on the mass of the produced system satisfying two conditions: 
$M_X < 25$ GeV and $0.01 \le x\le 1$ corresponding to the selection of 
approximately fixed length of the rapidity gap in experiment. 

\begin{figure}
\begin{center}
\epsfig{file=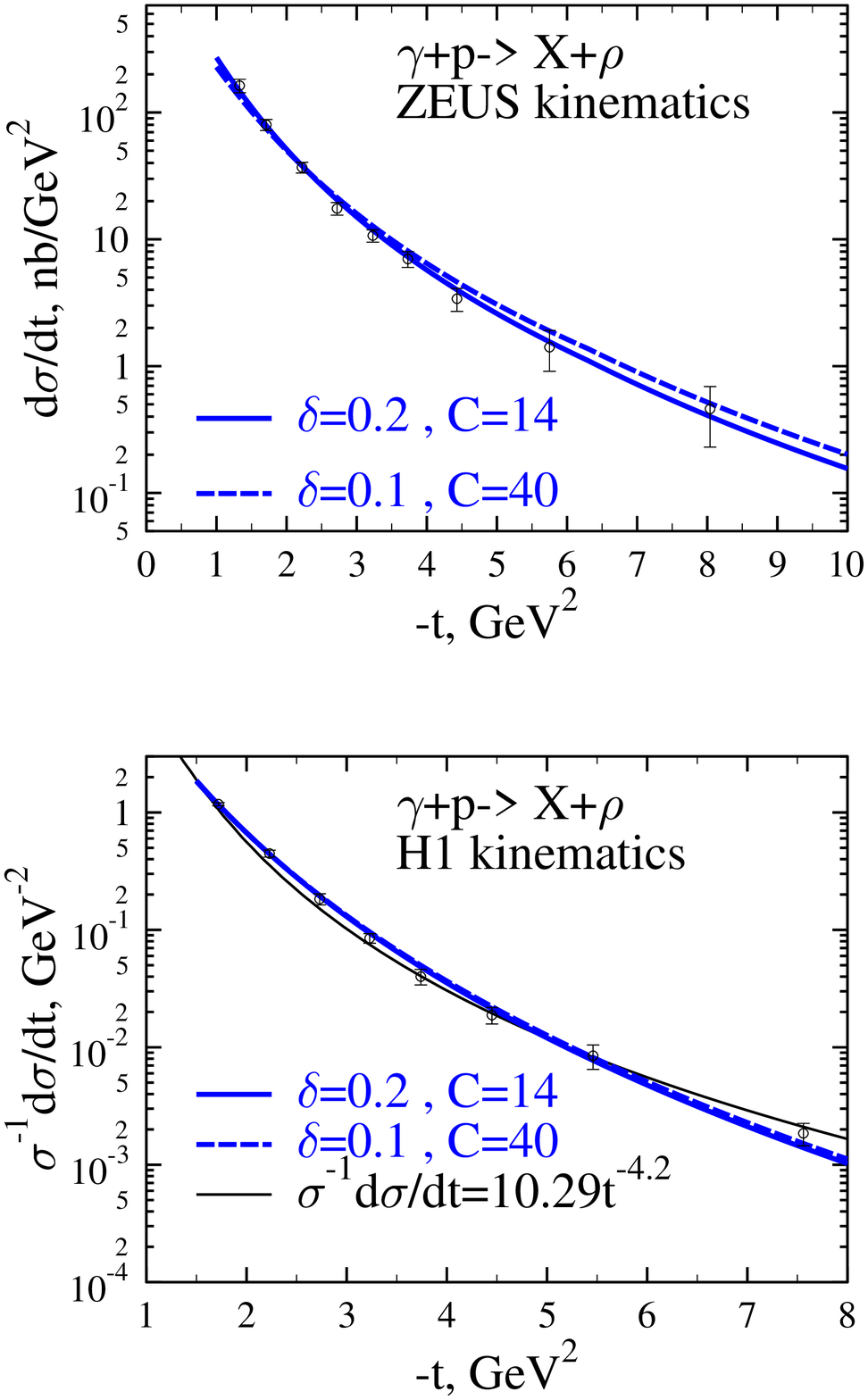, height=8in}
 \caption{ Description of ZEUS and H1 data for t-dependence of the large t 
and rapidity
gap cross section. ZEUS data were taken at average $W_{\gamma p}$=100 GeV 
with fixed cut
$M_{X}\le 25\, GeV$ and additional restriction $0.01 <x< 1$. The H1 data
were taken at average $W_{\gamma p}$=85 GeV and cut $M_{X}\le 5\, GeV$.}
 \label{zht}
\end{center}
\end{figure}

Based on the structure of  Eq.(\ref{DGLAPBFKL}) we  describe these HERA data 
using the following expression for the cross section 
\begin{equation}
{\frac {d\sigma_{\gamma +p\to \rho+X}} {dt}}=
\frac {C} { ({1-t/t_0} )^{4}}
 \biggl(\frac {s} {{m_V}^2 -t} \biggr)^{2\delta(t)}I(x_{min},t),
\label{basic}
\end{equation}
where
\begin{equation}
I(x_{min},t)=
\int \limits_{x_{min}}^{1} x^{2\delta(t)}
 \left[\frac {81} {16} g_{p}(x,t)+
\sum _{i}[q_{p}^i (x,t)+{\bar q}_{p}^i (x,t)\right]dx.
\label{intx}
\end{equation}
The gluon, quark and antiquark distributions were taken from the CTEQ6m PDF 
set\cite{cteq6} with account for up and down quarks and antiquarks. 
The value of $x_{min}$ was 
calculated using Eq.(\ref{xmin}) separately for each of experimental 
sets of the ZEUS and H1 data
using reported cuts on $M_X$. The scale factor $t_{0}=1\, GeV^2$ was fixed. 
The phenomenological function $\delta (t)$ which parameterizes the energy 
dependence of the the pQCD elastic amplitude
was chosen in the form $\delta (t)=\delta_{0}+{\delta}^{\prime}t$.
Values of $\delta_0 \,, \delta^{\prime}$ and the normalization constant  $C$
were adjusted to provide a reasonable  description of the 
available HERA data of the $\rho$ meson proton-dissociative
photoproduction\footnote{It is worth emphasizing 
here that the HERA data correspond to relatively small rapidity interval available 
for emission 
of the gluons in the color singlet ladder - $\ln (xs/Q^2) \le 5$. 
This  allows emission of one gluon 
in the ladder kinematics making it very difficult to apply a BFKL type approximation.}.
The t-dependence of cross section was measured 
by H1 and ZEUS for  different cuts on the produced mass $M_X$ and 
rather close values of the average energy $W_{\gamma p}$. 
As a result,  these data don't allow to fix unambigously energy dependence of the
dipole-parton amplitude (function  $\delta(t)$). We 
obtained a reasonable description of the data  assuming a relatively
weak energy dependence of the amplitude $\delta(t)=0.1$ ($C=40$)
as well as with more 
strong energy dependence $\delta(t)=0.2$ ($C=14$)  expected for
the hard processes (see Fig.\ref{zht}).  We want to emphasize here that these 
values of $\delta(t)$ are significantly larger that the ones which would result 
from extrapolation 
of the soft Pomeron trajectory to large $t$:
$\delta(t)=\alpha^{soft}_{\Pomeron}(t)-1\approx  0.08
+ t/4 \le 0_{\left|-t > 0.4 GeV^2\right.}$
even if one would introduce nonlinear term in the trajectory \cite{Erhan:1999gs}.

 Our results are consistent with a  rather weak variation of the energy 
dependence of the 
elastic amplitude with $t$ - we used the value  $\delta^{\prime}=0$ while the presence 
of very small negative value $\delta^{\prime}=-0.01$  will improve agreement with H1 
data at large $-t>5\,GeV^2$ in  the  case of calculations with $\delta_{0}=0.2$.
In difference from  the results of H1 and ZEUS analyses which found 
different t-dependence 
($|t|^{-3.2}$ in the ZEUS experiment and $|t|^{-4.2}$ in the H1 
experiment from their fits to the data
we are able to 
describe both data sets with the same universal expression and the 
same values of parameters.
This is due to the different $t$-behavior of $I(x_{min},|t|)$ 
in the kinematics of ZEUS and H1 because of different
cuts in experiments. In the first case the lower limit in integration
over $x$ is practically constant: $x_{min}\approx 0.01$ for all range 
$1.5 \,GeV^2 < -t < 8 \, GeV^2$ due to the condition $0.01< x<1$. 
In the H1 kinematics ($M_{X}<5\, GeV$) the value of integral
is decreasing with increase of $t$ because $x_{min}$ increases from
$x_{min}\approx 0.06$  to $x_{min}\approx 0.25$. 

As we mentioned above in the hard regime the energy dependence
of the amplitude should be a weak function of $t$.
 In photoproduction of $\rho $ meson with rapidity gap which  we consider here 
one needs large $t$ to 
reach dominance of the hard mechanism. However in the case of exclusive 
onium photo/electro production
or exclusive light vector meson electroproduction at large $q^2$ one expects 
the hard mechanism to dominate already at $t\sim 0$. 
Hence  the parameter $\delta$ should be close to the energy 
dependence of the amplitude of exclusive process $\gamma^* + p \to V + p$. At HERA 
the highest virtualities for such amplitude are reached  in the exclusive
$J/\psi$ electroproduction and  correspond to $\delta \sim 0.2$, for a 
summary see Ref. \cite{Levy:2005ap}. Hence our observation that 
similar value of $\delta$ allows to describe  the rapidity gap data at 
large t gives a support to the interpretation of the data as due to hard 
elastic quark-antiquark dipole - parton scattering.

\begin{figure}
\begin{center}
\epsfig{file=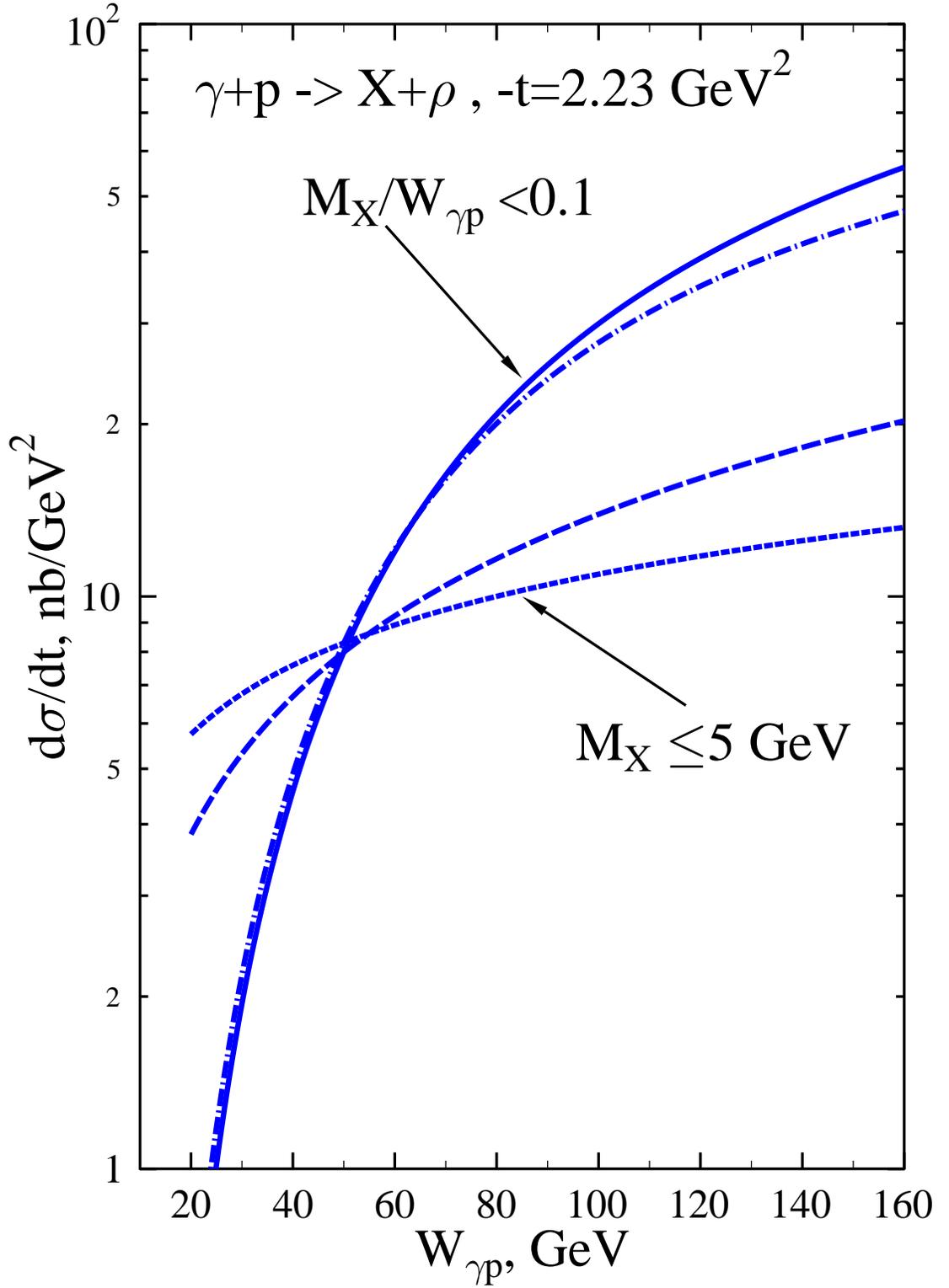, height=8in}
 \caption{Energy dependence of 
cross section given by Eqs. (\ref{basic},\ref{intx}) at $-t=2.23\,GeV^2$.
Solid and long dash lines - calculations with $\delta =0.2$ and $C=14$;
dot-dashed and short dash lines - calculations with $\delta =0.1$ and $C=40$. }
 \label{hw}
\end{center}
\end{figure}

\begin{figure}
\begin{center}
\epsfig{file=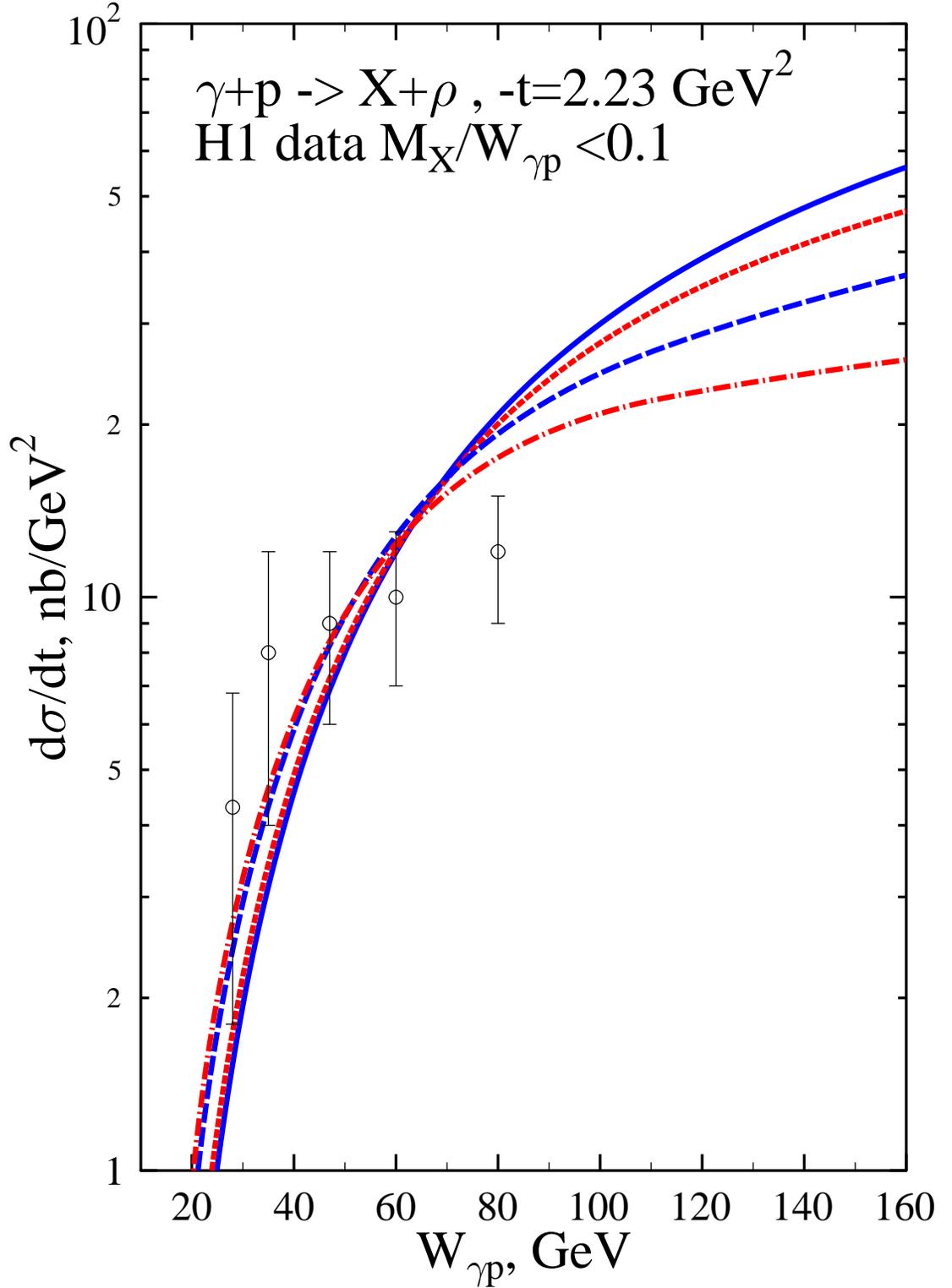, height=8in}
 \caption{Comparison of the preliminary H1 data \cite{Olsson}
  with energy dependence of 
cross section given by Eqs. (\ref{basic},\ref{intx}).
Solid and long dash lines - calculations with $\delta=0.2$ and virtuality of 
parton $Q_{eff}^2=-t$ and $Q_{eff}^2=-t/4$ correspondingly, short dash and 
dot-dashed lines - with $\delta=0.1$ and virtuality $Q_{eff}^2=-t$ and $Q_{eff}^2=-t/4$
correspondingly.}
 \label{hwt}
\end{center}
\end{figure}

Obviously, the parameter $\delta$ could be more reliably fixed 
analyzing the energy dependence of cross section for fixed values 
of $t$ in a wide range of $W_{\gamma p}$.  Such preliminary data 
in the range $20 \,GeV<W_{\gamma p}<100\, GeV$ were presented at 
DIS06 by H1 for $-t<2.5\,GeV^2$  for the cut
$M_{X}<0.1W_{\gamma p}$ \cite{Olsson}.  This experimental cut leads to 
decrease of $x_{min}$  from $x_{min}\approx 0.3$ 
to $x_{min}\approx 0.02$ with increase of $W_{\gamma p}$. 
Under these conditions Eq.(\ref{basic}) leads to  a strong increase of the cross 
section
due to the  strong growth of integral $I(x_{min},|t|)$
with increase of $W_{\gamma p}$ in the range from 20 GeV to 100 GeV. This 
 growth  
weakly depends on the  value of $\delta$ (Fig.\ref{hw}).
No results for the energy dependence of the process in
the kinematics where
 $M_X$ is restricted from above
 by some value $\hat M$ were reported by the HERA experiments so far. 
The reported 
energy dependence  is substantially weaker than the one given by Eq. (\ref{basic}) 
(Fig.\ref{hwt}). For  moderate $t$  the energy  dependence of the cross 
section is  sensitive to the scale of the gluon pdf's in 
Eq.\ref{DGLAPBFKL}. To illustrate this point we present in   Fig.\ref{hwt} the 
result of calculations 
for  two choices of scale in 
Eq.\ref{DGLAPBFKL}: $Q^2$ and ${Q^2}/4$.  It appears  that the energy 
dependence of the 
data is described better if the scale is $\le {Q^2}/4$.

\begin{figure}
\begin{center}
\epsfig{file=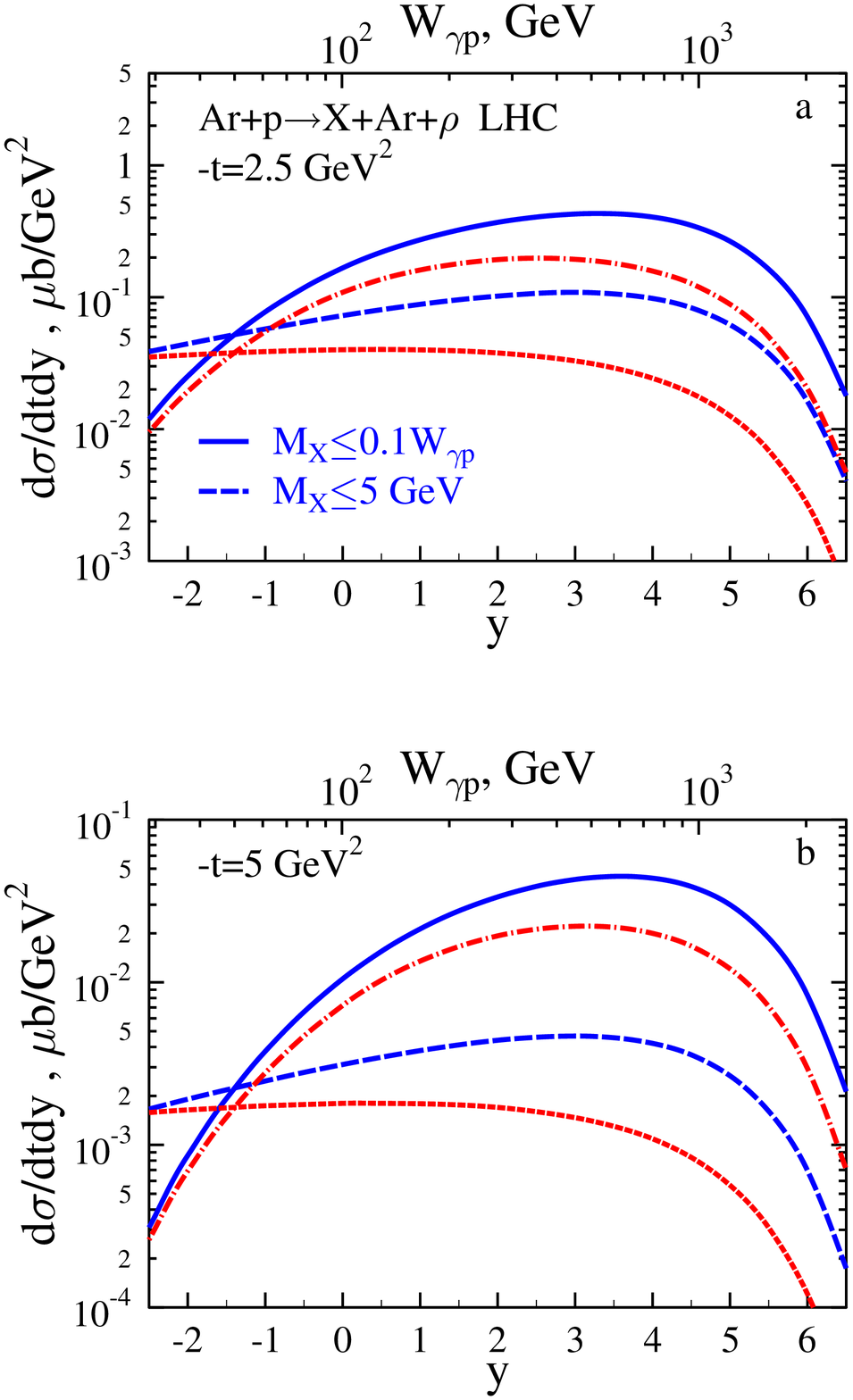, height=8in}
 \caption{Rapidity distribution for the large t and rapidity gap cross
 section of $\rho$ meson photoproduction in ultraperipheral proton-argon
 collisions in kinematics of LHC. Solid and dashed lines - calculations
with $\delta=0.2$, dot-dashed and short dash - with $\delta=0.1$.
The counting rates for pAr collisions can be estimated using the value of
luminosity $L=6 \mu b^{-1}sec^{-1}$ expected at LHC .}
 \label{arp}
\end{center}
\end{figure}

\begin{figure}
\begin{center}
\epsfig{file=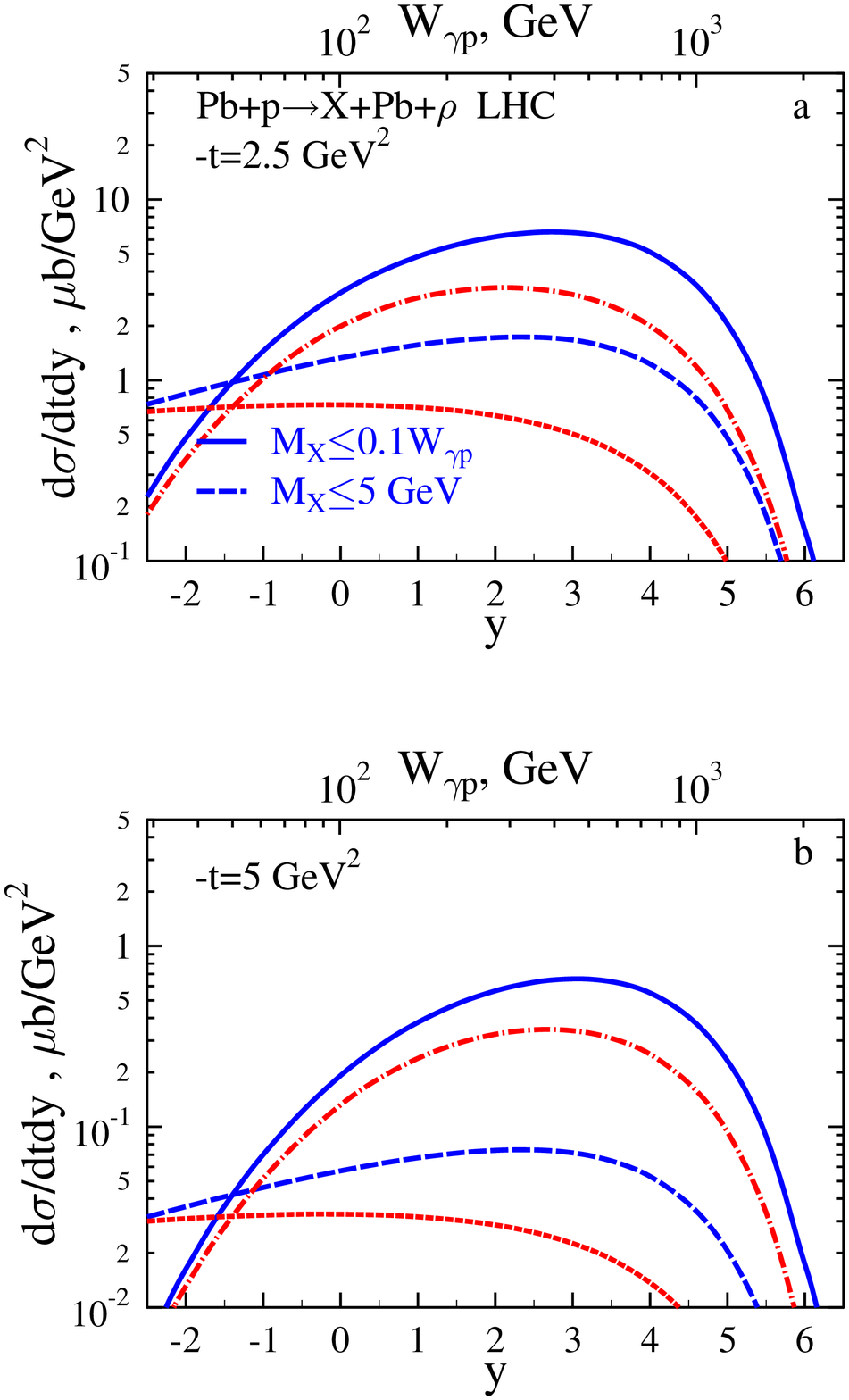, height=8in}
 \caption{Rapidity distribution for the large t and rapidity gap cross
 section of $\rho$ meson photoproduction in ultraperipheral proton-lead
 collisions in kinematics of LHC. Notation is the same as in previous figure.
The corresponding counting rates 
 can be easy estimated using the value of
luminosity of pPb colisions $L=10^{-1} \,\mu b^{-1}sec^{-1}$.}
 \label{pbp}
\end{center}
\end{figure}

It is not likely  that further studies at HERA will 
be able to cover a sufficiently wide 
 range of $W_{\gamma p}$ and rapidity gaps necessary to study the 
energy dependence of the large t elastic amplitude of "small size dipole" -  parton 
scattering. On the other hand at LHC CMS and ATLAS would have large enough 
rapidity coverage to observe reaction (\ref{eqvm}) 
in a wide range of $W_{\gamma p}$ and 
rapidity gaps in the ultraperipheral proton - nucleus collisions. Hence,   
we use parametrization of $\gamma +p\to X+\rho$ cross section given by
Eqs.\ref{basic}, \ref{intx} to estimate the large $t$ rapidity gap cross 
section of the 
$\rho$ meson photoproduction in the ultraperipheral proton-nucleus and  
(in the next section) in nucleus-nucleus collisions at LHC. For the case of 
proton-ion UPC 
we do not address here a small contribution  generated 
by $\gamma + A\to \rho + X $ reaction since it is much smaller 
and can be easily separated experimentally.
Hence, the large t nucleon dissociative cross section for 
this process will be given by expression
\begin{equation}
{\frac {d\sigma_{pA\to \rho X A}} {dydt}}=N_{\gamma}^{Z}(y)
{\frac {d\sigma_{\gamma N\to \rho X}(y)}{dt}},
\end{equation}
where $N_{\gamma}^{Z}(y)$ is the flux of photons
with energy $\omega=\frac {m_{\rho}} {2}e^{y}$ generated by fast
 moving ion with charge Z and 
the rapidity of vector meson is determined by expression
$y={\frac {1} {2}} \ln {\frac {E^V +p_L^V} {E^V -p_L^V}}$

We considered two regimes for the   $\rho$-meson
photoproduction with large rapidity gap at intermediate and large 
momentum transfer in UPC at the LHC analoguous to the ones 
studied at HERA: 
\begin{itemize}
\item 
The cross section can be studied at fixed t as a function of rapidity 
of the produced $\rho$ meson imposing the restriction $M_X\le 5 GeV$. 
In this case the lower limit $x_{min} $ of integration 
over $x$  does not depend on $W_{\gamma p}$
and the amplitude of dipole - parton elastic scattering 
varies with $W_{\gamma p}$ due to the  increase of the rapidity gap due to the increase of the  rapidity of the produced vector meson. Such approach could allow
one to study energy dependence of the dipole-parton amplitude and determine 
parameter $\delta$.
\item 
Cross section is studied for  ${M}_{X}=0.1\cdot W_{\gamma p}$. 
This type of cut is interesting since it 
corresponds to keeping  the rapidity gap fixed and changing the value 
of $x_{min}$. Such study could be useful for testing the parton distribution 
functions and the reaction mechanism by extracting from the data the
integral $I(x_{min},t)$ at different values of $x_{min}$ and $t$.

\end{itemize}

Note, that we do not consider here the photoproduction
cross section at $W_{\gamma p} < 20\,GeV$ where
our parametrization (Eqs. \ref{basic}, \ref{intx}) based
on the HERA data will be hardly reliable, in particular, for a case 
when the produced mass is fixed by the condition $M_{X}\le 5\,GeV$. 
On the other hand, one can see from Fig.\ref{hw}
that the cross section is very small if $M_X\le 2$GeV.

The results of calculations for $\rho$ meson nucleon dissociative 
photoproduction in proton-argon and proton-lead ultraperipheral
collisions at $-t=2.5\, GeV^{2}$ and $-t=5.0 \,GeV^{2}$ in the kinematics 
of the LHC are presented in Figs.\ref{arp},\ref{pbp}. Note here, that the 
cross section should drop rather slowly with t - the integrated 
rate for $-t> t_{min}$ drops approximately as $1/t_{min}^3$. 
One can see that with expected at the LHC luminosity of the proton-ion
collisions the rates remain high up to sufficiently large $t$. 
In particular, the rates for $-t > 10 \,GeV^2$ are only a factor of 10 
smaller than for $-t> 5 \,GeV^2$. 
It is also worth noticing that the rates for $J/\psi $ production 
would be also significant since though these rates for a fixed $t$ 
are smaller than for $\rho $ production, one would be able to use 
for the analysis $-t\ge 1 \,GeV^{2}$ where the discussed process 
gives much larger contribution than the exclusive diffractive 
process $\gamma +p\to J/\psi + p$.

The most of the events detected in the considered kinematics would correspond to 
$x\ge 0.01$, so the main information which will be possible to infer 
from the data would be on the energy dependence of the elastic 
$"quark-antiquark \,dipole"\, - \,parton $ amplitude for 
different virtualities. Some 
number of the events can also be collected for $x$ as low as $x\sim 10^{-3}$.
On the other hand it will be probably very difficult
to reach
 the $x\ge 0.4$-range  where scattering off quarks gives 
a larger contribution than scattering off gluons. 
 Overall it is clear that the energy range for the dipole - parton interactions 
will be large enough:  $s_{max}/s_{min} \ge 4\cdot 10^3$ to provide a precision 
 measurement of the energy dependence of the amplitude.
If $\delta\approx 0.2$, the elastic cross section should increase in this 
interval by a factor $\sim 30$.
 
\section{A-dependence of the cross section and estimates for
 the photoproduction in the heavy ion UPC.}
 \label{sec3gap}
 
  Since the large $t$ rapidity gap processes  $\gamma(\gamma^*) + N \to V + X$ are 
dominated by elastic scattering of small $q\bar q$ pairs off partons  in a 
target one can use these processes imbedded in nuclei to investigate in a novel 
way interaction of small dipoles with the nuclear medium.
 
 The  ultraperipheral nucleus-nucleus collisions 
at LHC will provide the first opportunity to investigate 
 physics of new QCD regime of high energy 
  strong interaction with small coupling constant  and large target thickness
(further studies will be possible at an electron-ion
collider).
The specific of nucleus-nucleus ultraperipheral collisions which differs them 
from the proton-nucleus collisions is that vector mesons are produced by photons 
emitted by both colliding nuclei. 
Hence, the cross section is given by the sum of two contributions 
\begin{equation}
{\frac {d\sigma_{AA\to \rho X AA^{\prime}}} {dydt}}=N_{\gamma}^{Z}(y)
{\frac {d\sigma_{\gamma +A\to \rho +X+A^{\prime}}(y)}
{dt}}
+N_{\gamma}^{Z}(-y)
{\frac {d\sigma_{\gamma +A\to \rho +X+A^{\prime}}(-y)} {dt}}.
\end{equation}
Here $\sigma_{\gamma +A\to \rho +X+A^{\prime}}$ is cross section 
of $\rho$ photoproduction off nucleus with mass number $A$, $X$ is
a product of the target nucleon diffractive dissociation and $A^{\prime}$
denotes a residual nucleus. Several neutrons will be produced due to electromagnetic excitation of $A'$ 
by the nucleus $A$ which emitted a photon involved in the reaction (\ref{eqvm}).

The system $X$ should  resemble the one produced in the deep inelastic 
scattering off a nucleus $A$ at similar x and $Q^2\sim -t$ 
(except that the proportion of the scattering off gluons and quarks is different). 
The spectrum of the hadrons 
 is given by a superposition of the processes 
of the fragmentation of gluons and quarks in proportion given by 
Eq.\ref{DGLAPBFKL}.  Also,  these hadrons should be balancing the 
transverse momentum of the vector meson.  Note here that the momenta 
of the leading hadrons in the rest frame of the nucleus are of the 
order  $-t/(2m_N x)$.
Hence, based on the measurements of the EMC 
\cite{EMC} we can expect that 
for  large $t$ and   for $x \le  5 \cdot 10^{-2}$ 
absorption effects for the leading hadrons are small.
Nevertheless, a few neutrons will be produced in the fragmentation region
of the nucleus due to the final state interactions of produced hadrons\cite{window}. Therefore in the discussed process either one 
or both Zero Degree Calorimeters would detect  several neutrons. 
Detection of hadrons in the system X allows to determine which of 
the nuclei generated  a photon, and hence determine invariant 
energy of the $\gamma A$ collision.

We already mentioned that study of  A-dependence of the process 
$\gamma +A\to \rho +X +A^{\prime}$ at large $t$ allows to reveal
the dynamics of the small  
dipole - nucleus  interaction.
Before discussing the expectations for A-dependence in this kinematics
let us estimate what one can expect for A-dependence in
the high energy nucleon dissociative $\rho$-meson 
photoproduction off nuclei at small $t$. 
At high energies
the process originates from the interaction of the photon in an average configuration 
which interacts inelastically with the strength comparable to that of 
$\rho$-meson. The fluctuations of the interaction strength in this case 
are rather small and
the photoproduction cross section
can be calculated using the Gribov-Glauber approximation
for high energy incoherent processes (note, 
that in quantum field theory the process is described by
non-planar diagrams, see discussion below):
\begin{equation}
{\frac {d\sigma_{\gamma+A\to \rho+X+A}} {dt}}=
A_{eff}{\frac {d\sigma_{\gamma+p\to \rho+X}} {dt}},
\label{iabr}
\end{equation}
where 
\begin{equation}
A_{eff}=2\pi \int \limits_{0}^{\infty} bdb 
\int \limits_{-\infty}^{\infty}dz\rho_{A}(b,z) 
\exp \biggl [-\sigma_{in}^{\rho N} \int \limits_{-\infty}^{\infty}  
\rho_{A}(b,z)dz \biggr ].
\label{aeffgap}
\end{equation}
Here $\rho_{A}(b,z)$ is the nuclear density ($\int d^3r\rho_A(r)=A$).
The effective number of nucleons $A_{eff}$ involved in the the process of the 
vector meson photoproduction off nucleus determines so called 
rapidity gap survival probability which can be estimated as
\begin{equation}
{\frac {A_{eff}} {A}}= 
{1\over A} \int d^2b  T(b) \exp(-\sigma_{in}^{\rho N} T(b)),
\label{aeffgapav}
\end{equation}
where 
$T(b) =\int^{\infty}_{-\infty}dz \rho_A(z,b)$ is the optical 
thickness.  In the range of high energies which could be reached in 
$\rho$ meson photoproduction off lead at LHC growth of the cross section 
$\sigma_{in}^{\rho N}$ is significant and
the resulting suppression becomes very large - 
$A_{eff}/A\sim A^{-{2\over 3}}$,
emphasizing peripheral character of the process.
 
  At large $t$ the dominant component of the photon responsible 
  for the nucleon dissociative 
photoproduction of vector mesons
off nucleus is the quark-antiquark pair characterized by
the small transverse size ($d\propto 1/\sqrt{|t|}$) - the
 $q\bar q$ dipole. So, with increase of $t$ leading and higher twist 
nuclear shadowing effects should decrease
due to color transparancy phenomenon. 
It has been understood long ago \cite{Mandelstam63,Gribovcom} 
that the contribution of planar (eikonal/Glauber rescattering) 
diagrams to the
amplitude for high energy collisions is cancelled in a quantum field
theory. Recently this result has been generalized to pQCD for the 
interaction of spatially small 
 dipole with a large size color singlet 
dipole via  exchange by two color octet amplitudes see \cite{BLV} or via
 exchanges by several color  singlet amplitudes\cite{Blok:2006ns}.
 Principal difference between description of scattering process within
quantum mechanics and QCD as the quantum field theory is that
with increase of the incident energy the 
 dipole with an increasing probability fluctuates into configurations with a   
larger and larger number of  constituents before the 
collision \cite{annual,Blok:2006ns}. Each of these constituents can 
interact 
only once  with a parton (dipole) of a target through an amplitude 
with vacuum quantum numbers in the crossed channel. Thus account
of causality and/or energy-momentum conservation leads to Gribov-Glauber
picture but for the interaction of projectile and target described 
as a collection of partons -  bare particles of QCD.

\begin{figure}
\begin{center}
\epsfig{file=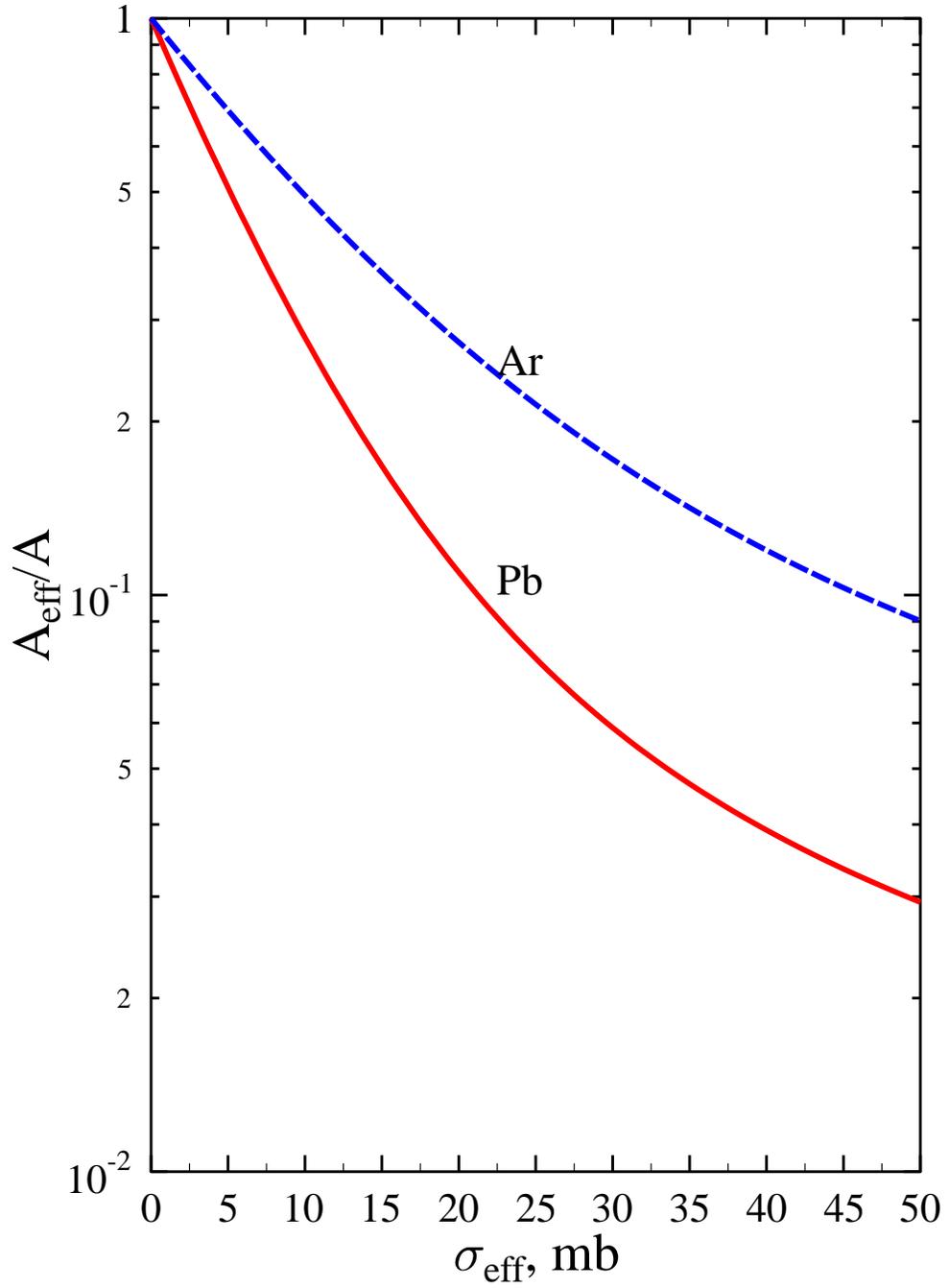, height=7in}
 \caption{The probability of rapidity gap survival as a function of 
$\sigma_{eff}$ which models the strength of dipole-nucleon interaction in
nuclear medium.}
 \label{gapsup}
\end{center}
\end{figure}

In the case of a small dipole - nucleus scattering 
 the first rescattering is given by pQCD cross section for the interaction 
of the $q\bar q$ dipole of the transverse size d with the nucleon which 
 in the 
leading order approximation can be written as \cite{sigma}
\begin{equation}
\sigma_{inel}^{q\bar q-N} (\tilde x, d^2) \;\; = \;\; 
\frac{\pi^2}{4} \; F^2 \; d^2 \; \alpha_s (Q^2_{\rm eff}) \;
\tilde x g_{T} (\tilde x, Q^2_{\rm eff}).
\label{sigma_d_DGLAP}
\end{equation}
Here $F^2 = 4/3$ is the Casimir operator of the fundamental
representation of the $SU(3)$ gauge group, $Q^2_{\rm eff}$ is the effective
virtuality and  $\tilde x=Q^2_{\rm eff}/W_{\gamma p}^2$.
Since the size of the dipole  scales as $1/\sqrt{|t|}$, at sufficiently large
$t$ and fixed $W_{\gamma p}$ the cross section $\sigma_{in}^{q\bar q-N}$
become so small that  
one can neglect interactions with 
$N\ge 3$ nucleons. Then the rapidity gap survival probability 
can be estimated using simple expression:
\begin{equation}
A_{eff}/A =1 - \sigma_{inel}^{q\bar q-N} {1\over A} \int d^2b T^2(b).
\label{first_order}
\end{equation}
 With increase of $W_{\gamma p}$ at fixed $t$ the dipole-nucleon
cross section $\sigma(q\bar q - N)$
rapidly  increases  due to the growth
of the small x gluon density
($\propto ({W_{\gamma p}^2/Q_{eff}^2})^{n}, n\ge 0.2$).
This results in breakdown of the Eq.\ref{first_order} and
one has to take into account
higher order rescatterings involving $N\ge 3$ nucleons which
interact with configurations containing three ($q\bar q g$)
and more partons. The cross sections of interaction with
such configurations should be larger than $\sigma_{inel}^{q\bar q-N}$
given by Eq.(\ref{sigma_d_DGLAP})
because the projectile
with a significant probability effectively  consists of several
dipoles of sizes comparable to the size of the initial dipole.
Hence the eikonal type expansion of the amplitude over the number of rescatterings
based on the averaged strength of interaction will obviously somewhat overestimate the
the absorption.
Still it appears reasonable to use the eikonal approximation  for the rough estimate of the  magnitude of the suppression.
In particular, in Fig.\ref{gapsup} we show calculation of $A_{eff}/A$
as a function of $\sigma_{eff}$ which one can consider
as a parameter modeling
the averaged
strength of dipole-nucleon interaction in nuclear  medium. The accuracy of such estimates of $A_{eff}/A$
should be better both in the limit of    small $\sigma_{eff}$
when $N > 2$ terms give a small correction
($\sigma_{eff} \le 3 \, mb$ for $A\sim 200$) and for large $\sigma_ {eff}$
where interaction is close to the BDR.

Clearly, increase of t at fixed energy leads to $A_{eff}/A\to 1$ revealing the onset of color transparency.
However for $W_{\gamma p}~sim 100 GeV$ typical for UPC collisions at  LHC and for upper range of HERA  a dipole of transverse size d=0.2 fm  (which may correspond to $-t \sim 10 \, GeV^2$) interacts with a  cross section
$\sigma_{in}^{q\bar q-N} \approx 5\, mb$ leading to $A_{eff}/A$ substantially smaller than one (see Fig. \ref{gapsup})\footnote{Note that for $ \sigma_{inel}^{q\bar q-N} \approx 5\, mb$, and  A=200,  Eq.(\ref{first_order}) leads  to $A_{eff}/A $ a factor of 1.6 smaller than 
 Eq.(\ref{aeffgapav})}.
Hence one probably would need a larger $t$, or smaller $W_{\gamma p}$ to
reach the regime of complete color transparency given by Eq.(\ref{ct})
  below.

Overall, we expect that at fixed $t$ increase of energy will gradually lead to the onset
of the regime of 
color opacity. 
At sufficiently high energies one is likely to reach a black disk regime (BDR)
for interaction of the dipole with the 
nuclear medium and the vector meson photoproduction would be 
strongly suppressed 
for central impact parameters,
leading to dominance of the peripheral process with the
cross section proportional to
$ A^{1/3}$. It is reasonable to assume that
suppression of the $\rho$-meson yield in this case would be comparable to the one 
in the soft regime which we discussed above and could be estimated 
  using  Eq.(\ref{aeffgapav}).
 
 Higher twist effects in this kinematics would be manifested also 
in the structure of the final state.  Due to a more peripheral nature 
of the higher twist mechanism large suppression of the A-dependence 
would be combined with a smaller break up of the nucleus which could 
be measured via study of the multiplicity of neutrons in the ZDC.

In the leading twist approximation
the cross section of vector meson photoproduction off nuclear target is 
given by Eq. \ref{DGLAPBFKL} where  parton distributions within a nucleon are substituted by nuclear parton 
density distributions  $g_A$, $q_A$ and $\bar{q}_A$ 
 \begin{equation}
{\frac{d\sigma_{\gamma + A\to V +X+A^{\prime}}} {dt dx}}
=
\frac {d\sigma_{\gamma+quark\to V+quark}} {dt}
{ [{81\over 16} xg_A(x,t) +\sum_i (xq_{A}(x,t)+x{\bar q_A}(x,t)) ]},
\label{ct}
\end{equation}
where  $g_A$, $q_A$ and $\bar{q}_A$ are the nuclear parton distributions. 
 
It is known experimentally that quark distributions for $0.05< x< 0.5$
do not deviate from the linear dependence by more than 10\%.
For the gluons (which dominate in Eq.~\ref{ct})  some current models predict an enhancement of up to 20\%  for  $x\sim 0.1$ which maybe followed by some suppression at $x\ge 0.4$.
Hence the leading twist approximation of 
Eq.~\ref{ct} leads to the prediction   
of the onset of color transparency regime 
with increase of $t$,
which is characterized by a strong suppression of the interaction of
small 
 dipole with nuclear medium. 
Then 
the upper limit of
the photoproduction cross section can be obtained in the Impulse
Approximation (IA)
\begin{equation}
{\frac {d\sigma_{\gamma+A\to \rho+X+A}} {dt}}=
A{\frac {d\sigma_{\gamma+p\to \rho+X}} {dt}}.
\label{ia}
\end{equation}

Note that studies of the discussed process at $x\ge 0.05$  allow 
to trigger on scattering of a high energy  photon in a small $q\bar q$ 
configuration without requiring that interaction involves scattering off 
the small x gluon field. This is in contrast with e.g. elastic photo/electro 
production of an onium state off nuclei.
 Accordingly this process provides a complementary clean 
way to study a pattern of the interactions of a small dipole with nuclear 
medium in a wide range of energies.

Of separate interest is the kinematics with $M_X$ corresponding to $x\le 0.01$
where one  expects a significant leading
 twist gluon shadowing of structure functions (the quark contribution in 
this kinematics is negligible) - on the scale of a factor of 
$\sim 2$ for $x\sim 10^{-3}, -t \sim 10 \mbox{GeV}^2$\cite{Guzey}. 
This should result in additional 
decrease of $A_{eff}/A$, hence, when $x$ becomes $\ll 10^{-2}$
 one would be   
addressing an interplay of the leading 
and higher twist effects in the dipole - parton interaction and 
the dipole - nuclear medium interactions.

 For numerical estimates of the rates of the discussed process at LHC we 
considered two scenarios - a color transparency regime  of cross section 
proportional to A - the impulse approximation
 and regime of strong absorption which can be considered as   a lower limit 
on the expected cross section. We made a natural assumption that absorption 
for the interaction of a small dipole for a given energy should not exceed 
the one for the interaction of a hadron build of same quarks. Accordingly 
we used Eq.(\ref{aeffgapav}) with 
 $\sigma_{in}^{\rho N}$ calculated basing on the fit of \cite{LD} 
for elastic $\rho$ meson photoproduction off proton and the vector dominance model.

Comparing  to a proton-nucleus  case two options of restricting  the mass 
$M_X$ of produced system were slightly modified to take into account correctly 
the low energy photoproduction that is necessary because of two-side
specifics of the symmetric nucleus-nucleus ultraperipheral collisions.  
So, we considered variation of the upper limit of $M_X$
with increase of the photon-nucleon
CM energy $W_{\gamma p}$ as $M_X \le 0.1W_{\gamma p}$.  As a
second option we used combined approach: $M_X \le 0.1W_{\gamma p}$
until $W_{\gamma p}\le 50\, GeV$
and fixed upper limit $M_X \le 5\,GeV$ at higher energies (central rapidities).
 
\begin{figure}
\begin{center}
\epsfig{file=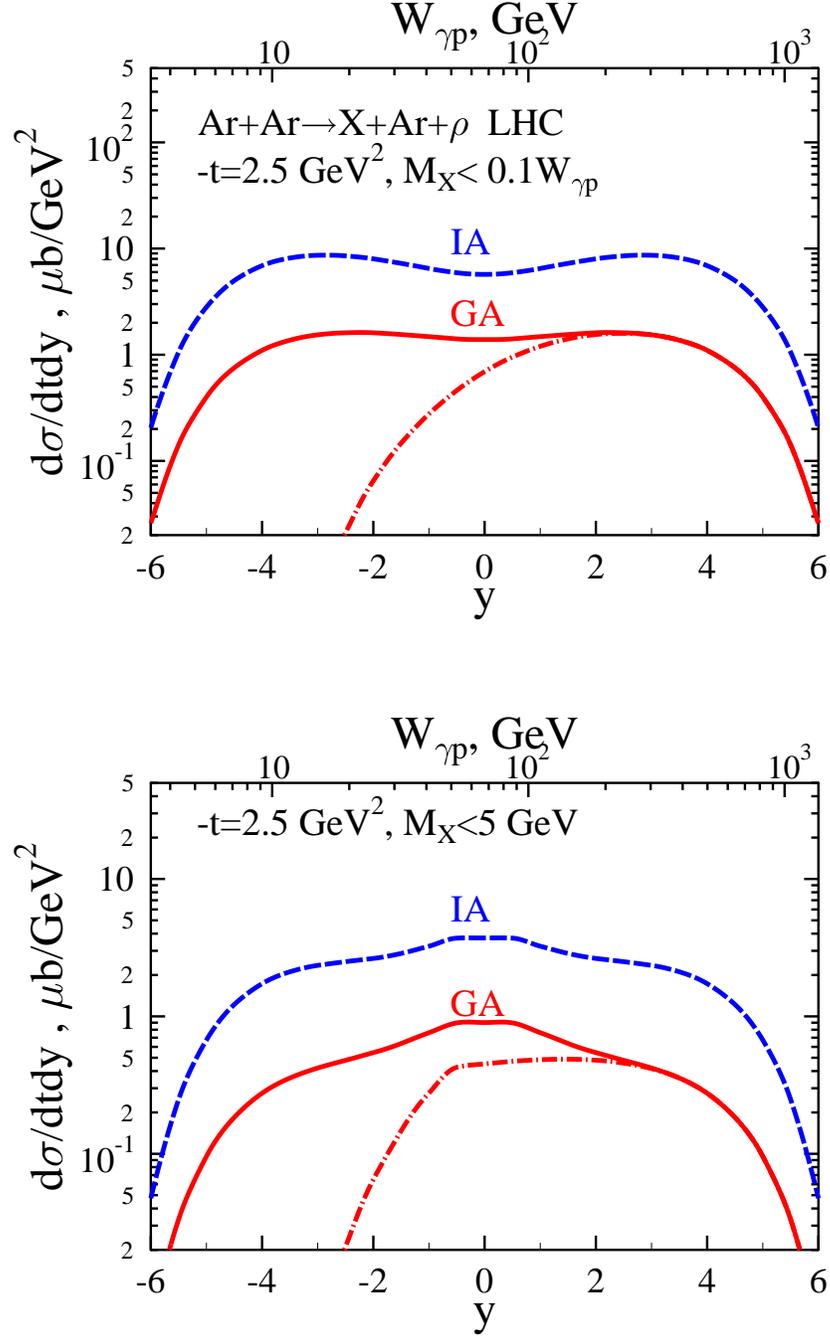, height=7in}
 \caption{Integrated over mass of produced system 
cross section of the nucleon dissociative
$\rho$ meson photoproduction at $-t=2.5\,GeV^2$ 
in the ultraperipheral argon-argon collisions at LHC. The upper 
figure - the limit of the mass of produced system
$M_X$ is proportional to the photon-nucleon center of mass energy
$M_{X}\le 0.1W_{\gamma p}$, in the lower figure for central rapidities
the limit of $M_X$ is fixed by restriction $M_{X}\le 5\,GeV$. 
Solid line - calculations with Glauber-Gribov screening, dashed line - calculations in the 
leading twist
approximation
neglecting nuclear shadowing correction which is very small for discussed kinematics, dot-dashed line - one-side contribution when $\rho$ meson
is produced by photons emitted by only one nucleus: large positive 
rapidities correspond to vector mesons produced by high energy photons. 
The counting rate can be estimated using expected luminosity for ArAr
collisions $L=1 \mu b^{-1} sec^{-1}$.}
 \label{art2i5}
\end{center}
\end{figure}

\begin{figure}
\begin{center}
\epsfig{file=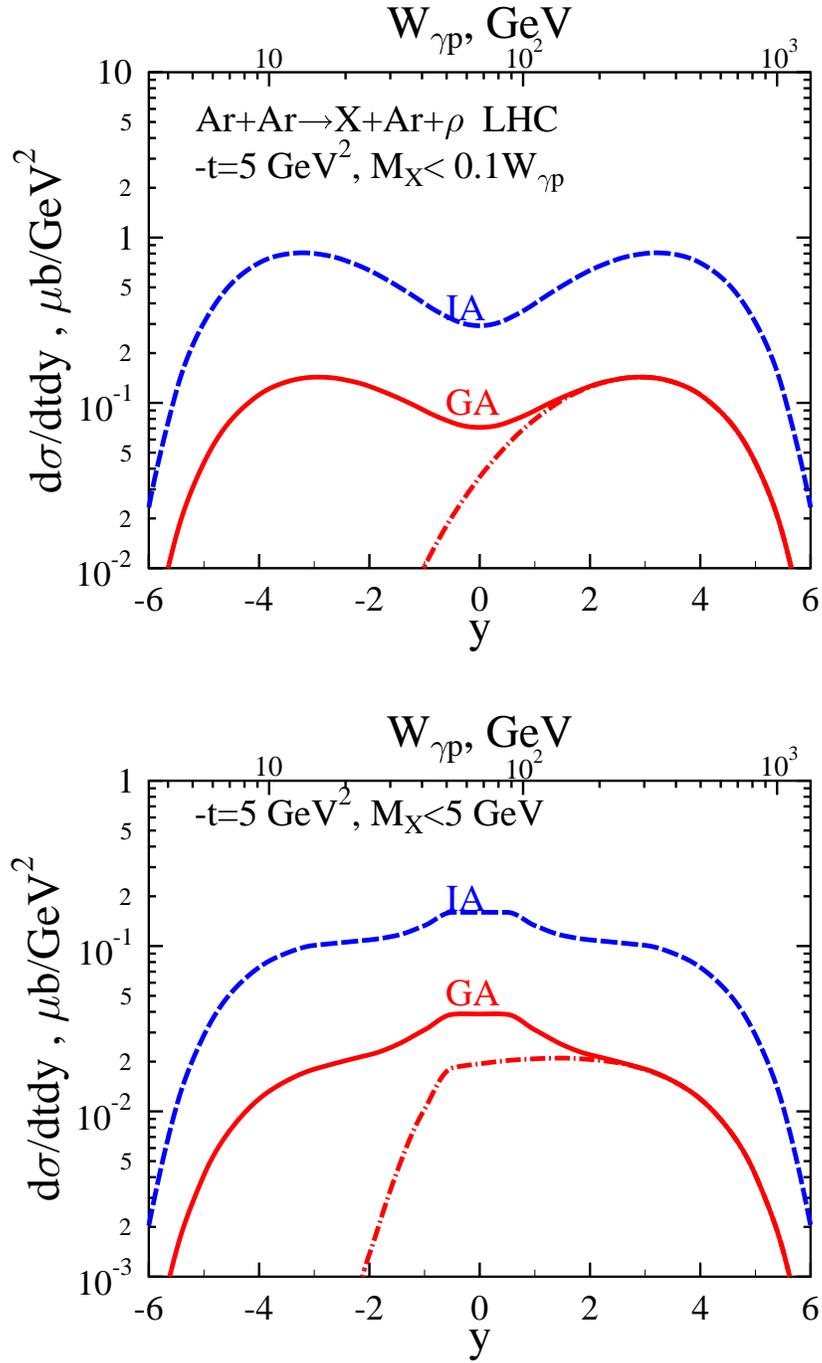, height=7in}
 \caption{The same as in Fig.\ref{art2i5} but 
at $-t=5\, GeV^2$}
 \label{art5}
\end{center}
\end{figure}

\begin{figure}
\begin{center}
\epsfig{file=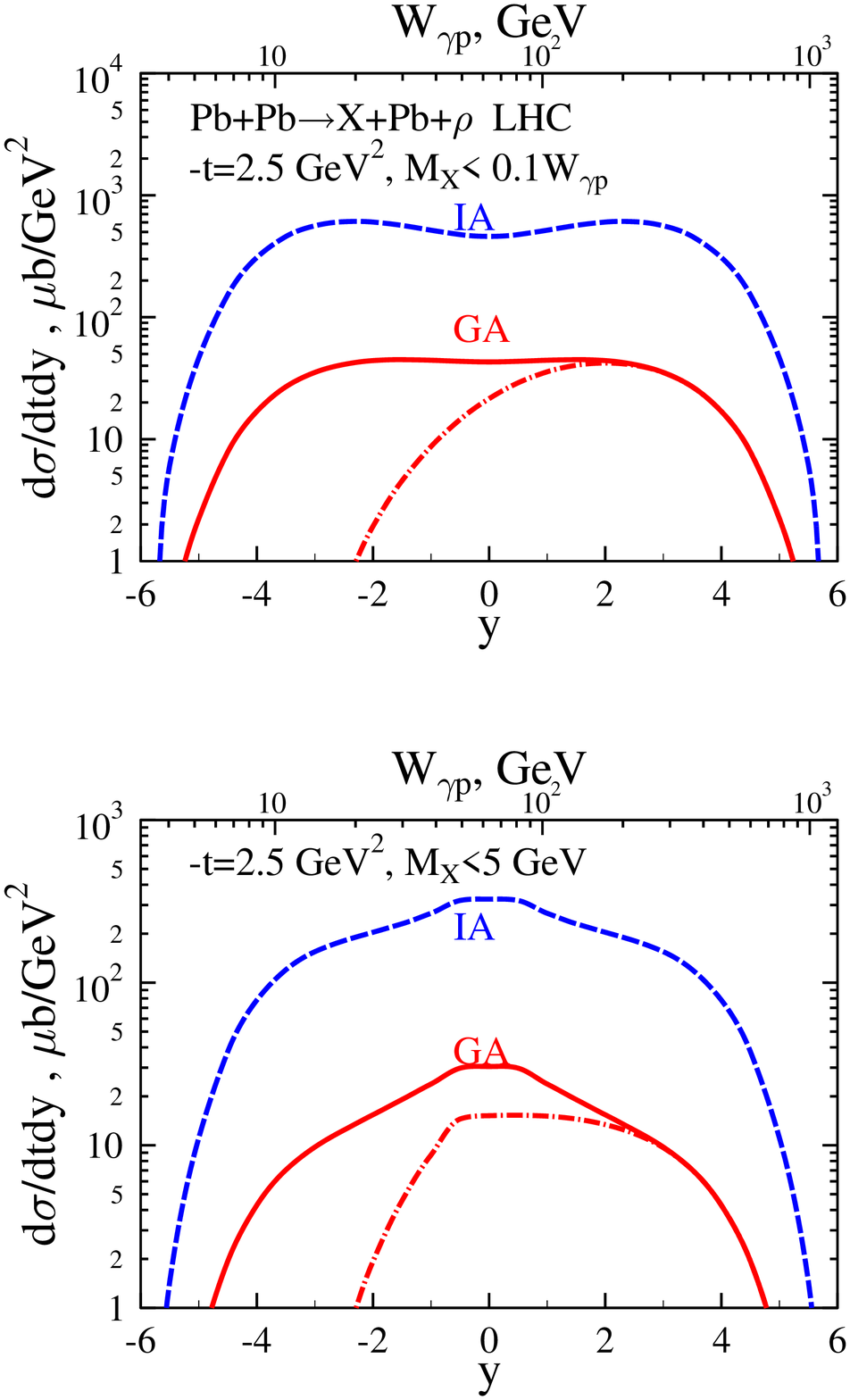, height=7in}
 \caption{
The same as in Fig.\ref{art2i5} but for PbPb UPC
and $-t=2.5\, GeV^2$. The counting rate can be estimated using
expected luminosity for PbPb collisions $L=10^{-3} \mu b^{-1}\,sec^{-1}$.
}
 \label{pbt2i5}
\end{center}
\end{figure}

\begin{figure}
\begin{center}
\epsfig{file=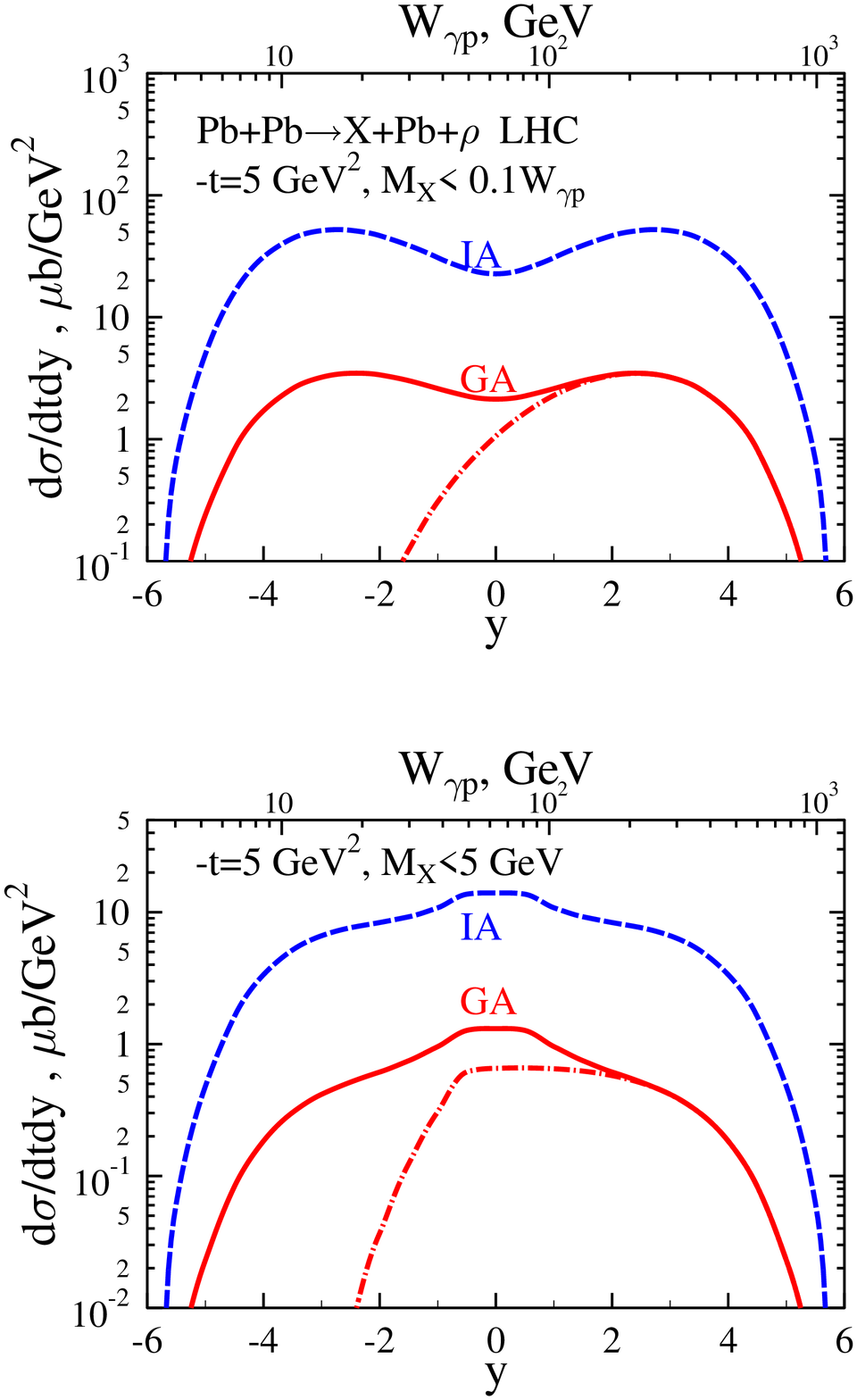, height=7in}
 \caption{The same as in Fig.\ref{art2i5} but for PbPb UPC 
and $-t=5\, GeV^2$}
 \label{pbt5}
\end{center}
\end{figure}

We present  the results of our calculations of the nucleon-dissociative $\rho$ 
meson photoproduction cross section  in ultraperipheral heavy ion 
collisions at LHC for a few values of the squared momentum transfer
in Figs. \ref{art2i5},\ref{art5},\ref{pbt2i5},\ref{pbt5}.

\begin{figure}
\begin{center}
\epsfig{file=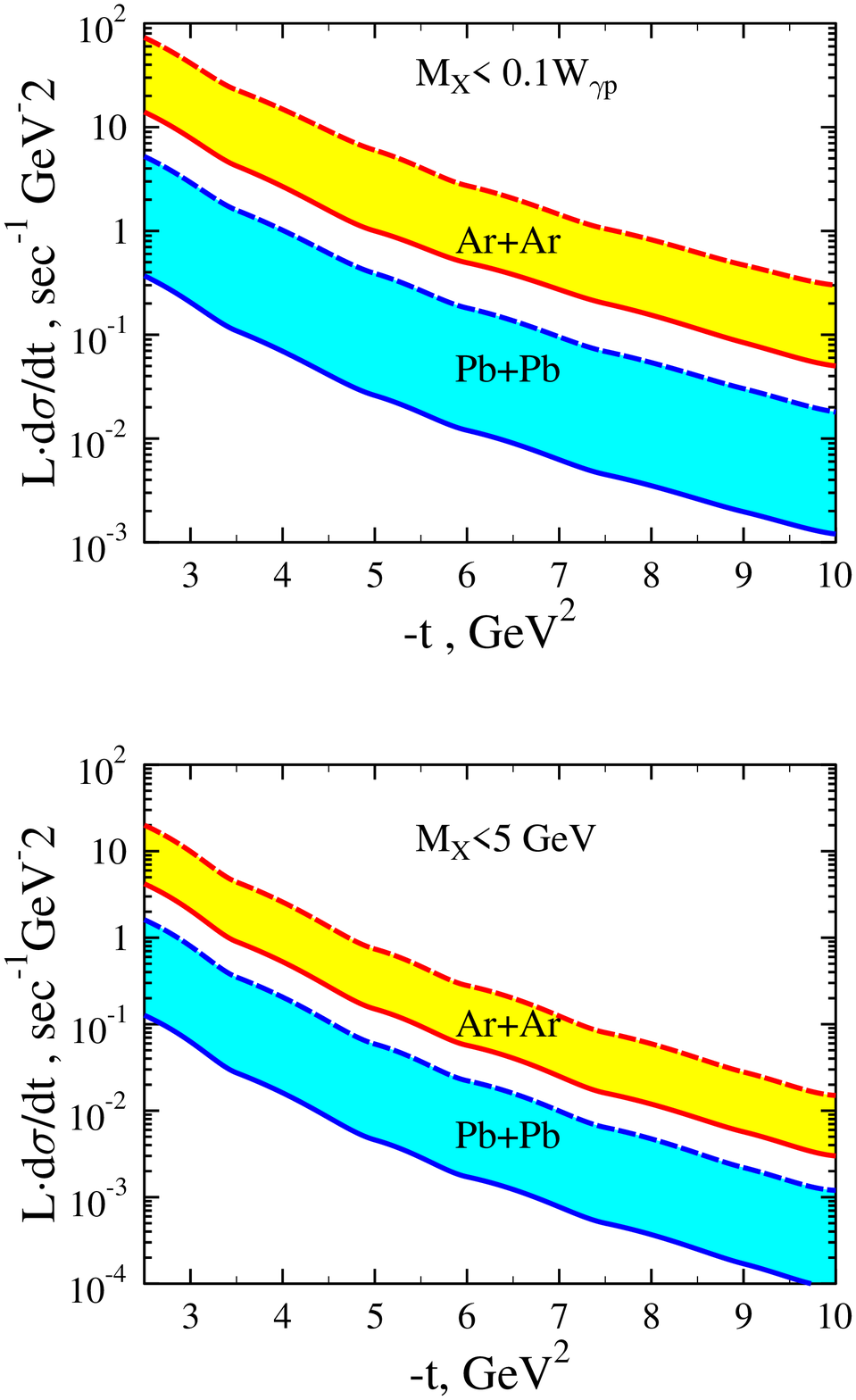, height=7in}
 \caption{Rapidity integrated 
counting rates for $\rho$ meson photoproduction with
rapidity gap in the UPC Ar+Ar and Pb+Pb collisions as a function
of $-t$. Lower bound of each colored space corresponds estimate
in the Gribov-Glauber approach, upper - in the Impulse Approximation.}
 \label{t10}
\end{center}
\end{figure}

One can see that for heavy nuclei 
suppression  may  reduce the cross section by a factor of ten and more. 
So the reduction depends strongly on the value of $q\bar q$ dipole interaction
with nucleons and,
hence, these processes are very sensitive to higher twist effects.
Even in the case of large screening effects, the rates would be large enough
up to $-t\sim 10 GeV^2$ (see Fig.\ref{t10}) and it 
would be doable to  study how the screening effects 
are reduced with increase of $t$.  This information would  be 
complementary to the information about energy dependence of the 
elastic small dipole - parton amplitude which is studied in the 
scattering off protons.

 \section{Conclusions}
 
To summarize, we have have found that the HERA data suggest a much faster  
energy dependence of the dipole - parton amplitude than in the soft dynamics. 
However the cuts used by the experiments make it difficult a direct determination 
of this energy dependence. We demonstrated that the study of the processes 
with rapidity gaps being feasible in UPC at LHC  will allow a direct measurement 
of important property of the pQCD - energy dependence of the large t elastic 
amplitude of dipole-parton scattering. In the case of $\rho$ meson production off 
nuclei we argued that there exist several different regimes of A-dependences 
related to the transition from a soft regime to the color transparency regime with increase of $-t$ for fixed $W$ and from 
color transparency to color opacity regimes for fixed 
$t$ and increase of $W_{\gamma p}$, to an 
interplay of the leading and higher twist effects  which is a nontrivial 
function of the rapidity gap interval. 
All together this provides a new powerful tool for  studying the 
small dipole interaction with the media.

\end{document}